\documentclass[journal=langd5,manuscript=article]{achemso}
\usepackage[version=3]{mhchem} 
\usepackage{color}
\author{Felix Gerlach}
\email{gerlach@sla.tu-darmstadt.de}
\phone{+49 (0)6151 1622198}
\affiliation[TU Darmstadt]
{Institute for Fluid Mechanics and Aerodynamics, Technische Universit\"at Darmstadt, Darmstadt, Germany}
\author{Maximilian Hartmann}
\affiliation[NMF]{Institute for Nano- and Microfluidics, Technische Universit\"at Darmstadt, Darmstadt, Germany}
\author{Cameron Tropea}
\email{ctropea@sla.tu-darmstadt.de}
\affiliation[TU Darmstadt]
{Institute for Fluid Mechanics and Aerodynamics, Technische Universit\"at Darmstadt, Darmstadt, Germany}

\title[Corner interactions during wetting]
  {The interaction of inner and outer surface corners during spontaneous wetting}
\keywords{Wetting, Rivulet, Cusp, Interaction}

\begin{document}

\begin{tocentry}

\includegraphics[height=3.5cm]{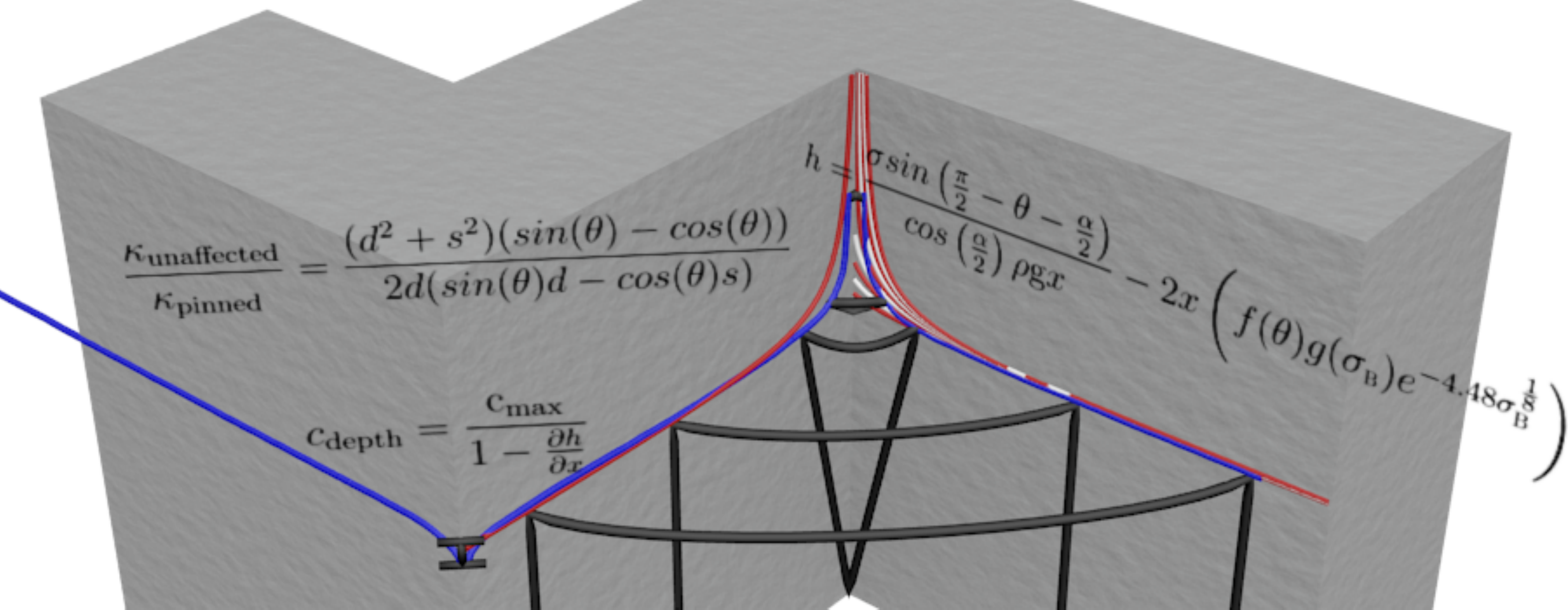}





\end{tocentry}

\begin{abstract}
Real world surfaces can often be modeled as a collection of edges, corners, dents or spikes of varying roundness. These features exhibit  individual spontaneous wetting behaviors comprising pinned contact lines, rivulets or cusps. If occurring in proximity to one another, as is often the case in applications, these wetting properties interact, resulting in an overall changed wetting pattern on the surface.  Hence, there is considerable interest in understanding when, and to what extent, interactions occur, and how wetting then deviates from the wetting of isolated surface features. 

The present study addresses these questions by experimentally and theoretically studying the capillary interaction of sharp-edged 90$^{\circ}$ (outer) and 270$^{\circ}$ (inner) corners in proximity to one another. It is shown that the spontaneous wetting at the convex outer corner is influenced by the concave inner corner even when they are separated by a distance of several times the capillary length, while the wetting of the inner corner takes place unaffected by the outer corner, except when the separating distance is much smaller than the capillary length. The final contact line shape at the inner corner is measured and theoretically modelled for contact angles up to 90$^{\circ}$.
\end{abstract}

\section{Introduction}
Many wetting applications exhibit geometrically complex surfaces. These surfaces can be either intended, like in gravure printing \cite{Hewson-2011-ID152, Schneider-2019-ID308} or exterior water management on vehicles \cite{Jilesen-2015-ID310, Dianat-2017-ID311}, or unintended, like rough casted metal parts or other inherent manufacturing features or complex assemblies, which then have to be retroactively coated or treated in some way to establish a desired surface finish. Hence a physical and theoretical understanding of wetting phenomena induced by macroscopic surface features is important to optimize such processes.

A  geometry commonly studied over the last 50 years are vertical cylindrical rods with variable spacing \cite{Princen-1969-ID288, Clanet-2002-ID206, Ponomarenko-2011-ID202}. From this geometry a smooth transition exists, starting from fibers \cite{Hsieh-1992-ID303, Quere-1988-ID297}, to natural fibers \cite{Hsieh-1992-ID303, Fuentes-2011-ID299}, to bundles of fibers \cite{Aranberri-Askargorta-2003-ID298}, towards fabrics \cite{Hsieh-1992-ID303, Hsieh-1996-ID301, Kim-2001-ID300, Yanilmaz-2012-ID302}. This transition shows how a complex real world surface can be constructed and described by the combination of simple geometric elements. Despite the round shape of the vertical cylindrical bodies, sharp corners are formed at their contact points, representing a primary influence on wetting. Hence, it is of major importance to understand the wetting at sharp vertical corners to understand the wetting physics of both simple and more complex surfaces.

For natural fibers and fabrics, in additional to the macroscopic rod shape of the fibers, micro- and nanometer features affect the wetting.  The effects of such micro- and nanoscale surface features are often modelled using pillar structures or general roughness elements \cite{Bico-2001-ID209, Palasantzas-2001-ID307, Boreyko-2011-ID210, Papadopoulos-2012-ID208}, using modified parameters such as the macroscopic contact angle or contact angle hysteresis, without changing the wetting phenomenon of the contact line itself. On the other hand, nanodefects \cite{Giacomello-2016-ID280} and anisotropic features such as groove-like structures \cite{Rye-1996-ID305, Yost-1997-ID304, Seemann-2005-ID306, Herminghaus-2008-ID3, Deng-2014-ID271} can trigger pinning, locally manipulating the contact line. Pinning defects obscure other wetting phenomena and  are therefore circumvented in the current study by using highly wetting silicone oil as a working fluid.

Often, when reporting on the wetting of rod-like elements in literature the liquid rise height is only given pointwise or as an average value. However, in the current study the entire contact line contour is examined. From this additional data, more information about the physics at, and the interaction between, different features (in this case corners) can be obtained. For instance the curvature of liquid close to an inner corner is described theoretically in the literature \cite{O'Brien-1968-ID293, Tang-1994-ID240, Ponomarenko-2011-ID202}, but only qualitatively or pointwise compared to experimental results, due to the lack of quantitative data. In order to obtain quantitative data over the entire contact line, flat walled samples are used in the current study to allow the calculation of the actual  contact line contour. Moreover, many real world surfaces can be modeled using collections of flat walls with different sharp or rounded corners in-between. This modelling works in the same manner as fabrics being modelled as a collection of cylindrical fibers. Using this analogy, the  geometry  in the current study, consisting of flat walls connected by two sharp corners, can be utilized to improve the understanding of the wetting of this family of complex surfaces by studying their simplest base geometry.

Additionally to the study of the contact line shape, the liquid in the corner is tracked in a similar way to the study of Ponomarenko et al. \cite{Ponomarenko-2011-ID202}, not analysing the late, but the initial phase of the liquid rise, which is not considered in many previous studies about wetting of corners.

\section{Experimental Details}
\begin{figure}
\centering
\includegraphics[height=0.3\textheight]{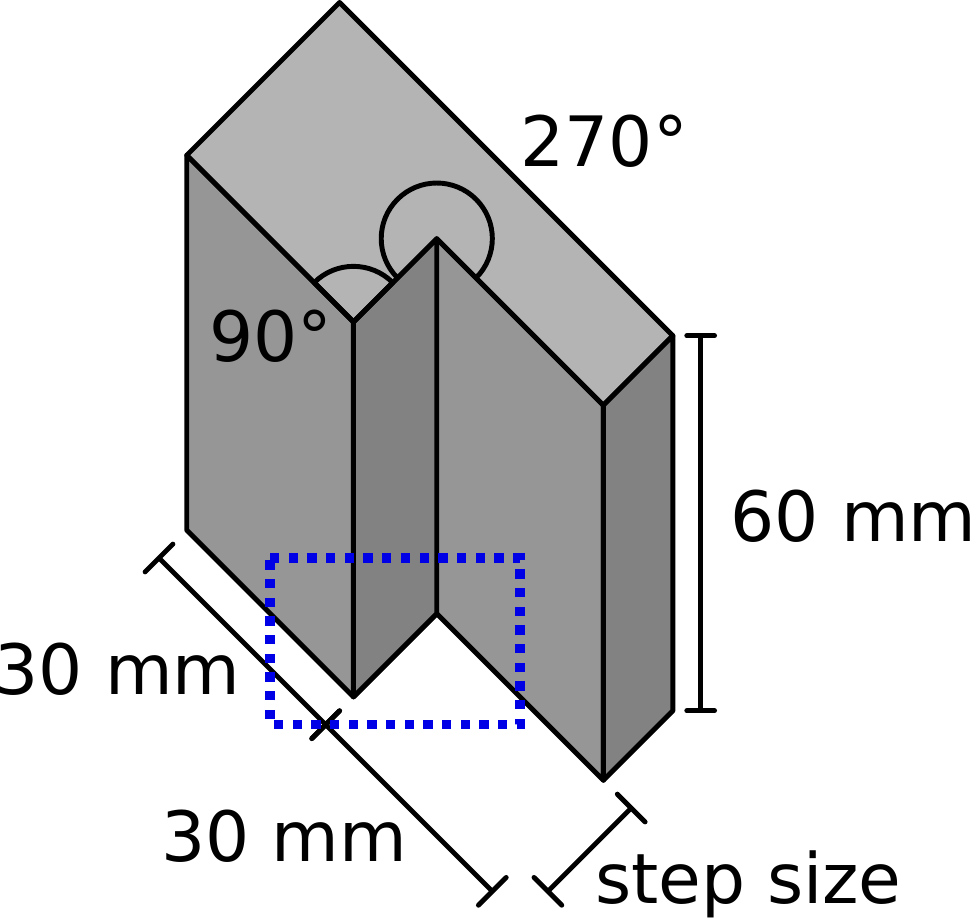}
\caption{The geometry of the samples used in this study.}
\label{fgr:sample}
\end{figure}

The current study uses samples made from aluminium alloy 2024 (AlCuMg2), milled to two different levels, creating a vertical step with a 90$^{\circ}$ and a 270$^{\circ}$ corner between the two flat faces, see Figure \ref{fgr:sample}. The length called \textbf{step size} in Figure \ref{fgr:sample} is a geometrical parameter which was varied during the experiments by using different samples. Figure \ref{fgr:sample} shows all relevant lengths for the results of the experiments. All dimensions of the samples, except that of the step height, which was varied according to Table \ref{tbl:steps}, are much longer than the capillary length, so that other parts of the sample do not influence the wetting at the two corners in the middle of the sample.

Since the current study focuses on the effects of macroscopic surface features, silicone oil (Polydimethylsiloxane, brand: ELBESIL SILIKON\"OL B) is used as wetting liquid, because its low surface tension ($\sigma = 20.6$ $\frac{\text{mN}}{\text{m}}$) and contact angle ($\theta < 20^{\circ}$) prevent almost all pinning effects. It has a dynamic viscosity  $\mu$ of 20 cSt and a capillary length l$_{\sigma}$ ($=\sqrt{\frac{\sigma}{\rho g}}$) of $\sim$1.48 mm, calculated using the liquid density $\rho$ and the acceleration due to gravity $g$. 

\begin{table}
  \caption{Analyzed step sizes.}
  \label{tbl:steps}
  \begin{tabular}{c|c}
    \hline
    Step size [mm] & Step size [l$_{\sigma}$]\\
    \hline
    0.05 & $\sim$0.033\\
    0.1 & $\sim$0.066\\
    0.25 & $\sim$0.166\\
    0.5 & $\sim$0.33\\
    1.5 & $\sim$1\\
    5 & $\sim$3.3\\
    10 & $\sim$6.6\\
    15 & $\sim$10\\
    \hline
  \end{tabular}
\end{table}

The sample is dipped into bulk liquid using a linear motor (PI LMS-180 408). This motor is assisted by  a low friction pneumatic cylinder (AirpotCorp AirpelPlus MP16S420NX) to counterbalance the  weight of the linear motor carriage and sample mount. The samples are dipped with a speed of 0.01 mm/s to avoid inertial effects, leading to a spontaneous liquid rise on the surface of the samples. During dipping the samples are illuminated with white light from the bottom front upwards while a B/W camera (pco edge 5.5) films them from the top front downwards. The light from the light source (Constellation 120 E) is directly reflected into the camera by the sample, while the rising liquid near the contact line refracts the light, leading to a visible shadow in the image. Although this illumination works well for a large area of the sample, the contact line is difficult to discern in some areas and has to be distinguished from light reflections due to the refraction of the liquid and the texture of the sample surface itself (Figure \ref{fgr:contactLine}). Hence, an automatic detection of the contact line is error-prone and manual contact line detection is used in the current study. Subjectivity in this manual procedure is partly avoided by repeating and averaging most measurements by several people.

To receive quantitative data regarding the contact line contour from the camera perspective, a post-processing algorithm is used to project the camera image onto a virtual model of the surface, which yields the three-dimensional position of the contact line.

\section{Results and Discussion}

\subsection{General phenomena at the two corners}

\begin{figure}
\centering
\reflectbox{\includegraphics[width=\textwidth]{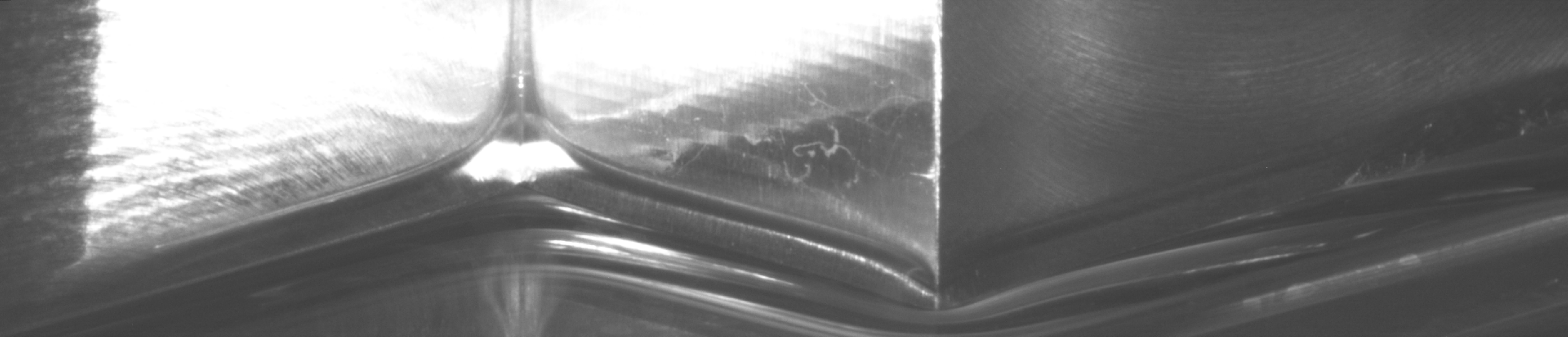}}
\caption{The final contact line at a 15 mm step.}
\label{fgr:contactLine}
\end{figure}

Figure \ref{fgr:contactLine} shows the final contact line on a 15 mm ($\sim$10 l$_{\sigma}$) step, filmed inside the dotted  frame sketched in Figure \ref{fgr:sample}. The sample in this experiment was mounted horizontally mirrored compared to the sketch in Figure \ref{fgr:sample}. To simplify the image comparison, the camera image in Figure \ref{fgr:contactLine} is shown mirrored. While the upper part of Figure \ref{fgr:contactLine} shows the three front faces of the sample, the triangle in the bottom left and the triangle at the bottom of the right image side show the transparent liquid pool surface through which the wetted part of the sample and other refractions and reflections can be seen. Between the dry sample faces and the liquid pool, the mostly dark liquid meniscus can be observed at the bottom of the sample. Due to light refraction, the meniscus shows some white lines and a bright area in the 270$^{\circ}$ back corner. Even after some image enhancement the contrast at the left sample face remains low.

The contact line appears oblique in the photograph, but when comparing the camera image with the perspective sketch from Figure \ref{fgr:sample} it becomes evident that this pattern and the triangular shape of the contact line contour result from the perspective view and represent horizontal lines along the sample. It is expected from the well understood wetting of vertical walls to see a horizontal contact line at the top of a static meniscus. While this basic behavior can be observed on the flat walls, the contact line deviatesis no longer straight close to the two corners of the sample. At the inner 270$^{\circ}$ corner (right) the contact line rises nearly vertically, which is called rivulet in this context. At the outer 90$^{\circ}$ corner (left) a depression of the contact line can be seen, which is referred to as a cusp.

\begin{figure}
\centering
\includegraphics[width=0.75\textwidth]{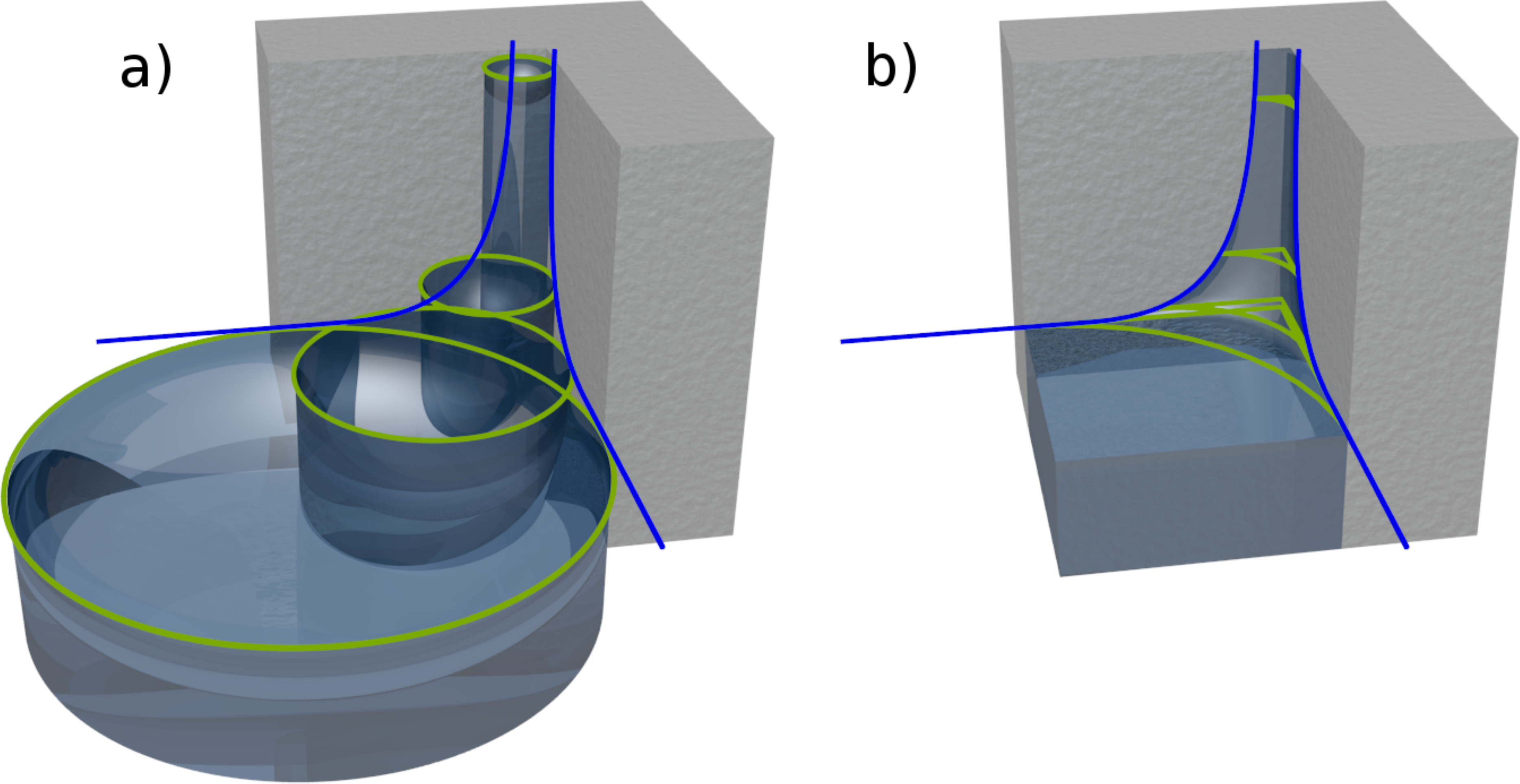}
\caption{Two explanations for the rivulet. a) The corner is seen as a set of round capillaries, fitting between the walls, which rise heights correspond to the local rivulet rise height\cite{Ponomarenko-2011-ID202}. b) Horizontal liquid curvature leading to a negative capillary pressure pulling up the liquid\cite{Tang-1994-ID240}.}
\label{fgr:rivuletExplanation}
\end{figure}

The formation of a rivulet in the inner corner can be explained using two models which are visualized in Figure \ref{fgr:rivuletExplanation}. The first model shown in Figure \ref{fgr:rivuletExplanation} a) is used by Ponomarenko et al.\cite{Ponomarenko-2011-ID202} and assumes the rivulet to behave at every point like liquid in a capillary tube fitting into the given space between the walls (green circles). Due to the decreasing size of the capillary tubes towards the corner the capillary rise height increases (due to a lower Laplace pressure), forming a rivulet of infinite height at the corner (blue lines). While this theory considers only the vertical curvature of the liquid, the theory shown in Figure \ref{fgr:rivuletExplanation} b) used by Tang et al.\cite{Tang-1994-ID240} considers only the horizontal curvature of the liquid and assumes the liquid to build horizontal arcs between the walls (green lines) which induce a negative capillary pressure at the corner, letting the liquid rise. Due to the decrease of arc radius with increasing height the pressure difference increases, also leading to an infinite rise in the corner (blue lines).

Concus et al. \cite{Concus-1969-ID294} showed that an infinite rivulet rise can only happen if the contact angle of the liquid ($\theta$) added to half of the opening angle of the given corner ($\tfrac{\alpha}{2}$) is below 90$^{\circ}$ ($\theta+\tfrac{\alpha}{2} < 90^{\circ}$). Since the inner corner opening angle in the current study is 90$^{\circ}$, an infinite rivulet rise should only occur for contact angles below 45$^{\circ}$, which was numerically confirmed by Thammanna Gurumurthy et al. \cite{ThammannaGurumurthy-2018-ID232}.

This behavior could be observed in preliminary experiments when monoethylene glycol was used as a liquid, which, depending on the  cleaning method for the samples, exhibited contact angles above 45$^{\circ}$ in some experiments. For water in combination with dirty samples an infinite rivulet rise could no  longer be observed. However, due to the strong pinning of these two liquids, they were not used for the final experiments discussed here.

Both models from Fig.~\ref{fgr:rivuletExplanation} do not consider the so-called Concus-Finn criteria. While the model from Fig.~\ref{fgr:rivuletExplanation}a) predicts an infinite rivulet rise for every contact angle above 90$^{\circ}$ due to the infinitely small capillary directly in the corner, the model from Fig.~\ref{fgr:rivuletExplanation}b) switches its limit from +$\infty$ to -$\infty$ at a threshold of 45$^{\circ}$, at which point it predicts a flat liquid surface.

\begin{figure}
\centering
\includegraphics[width=0.3375\textwidth]{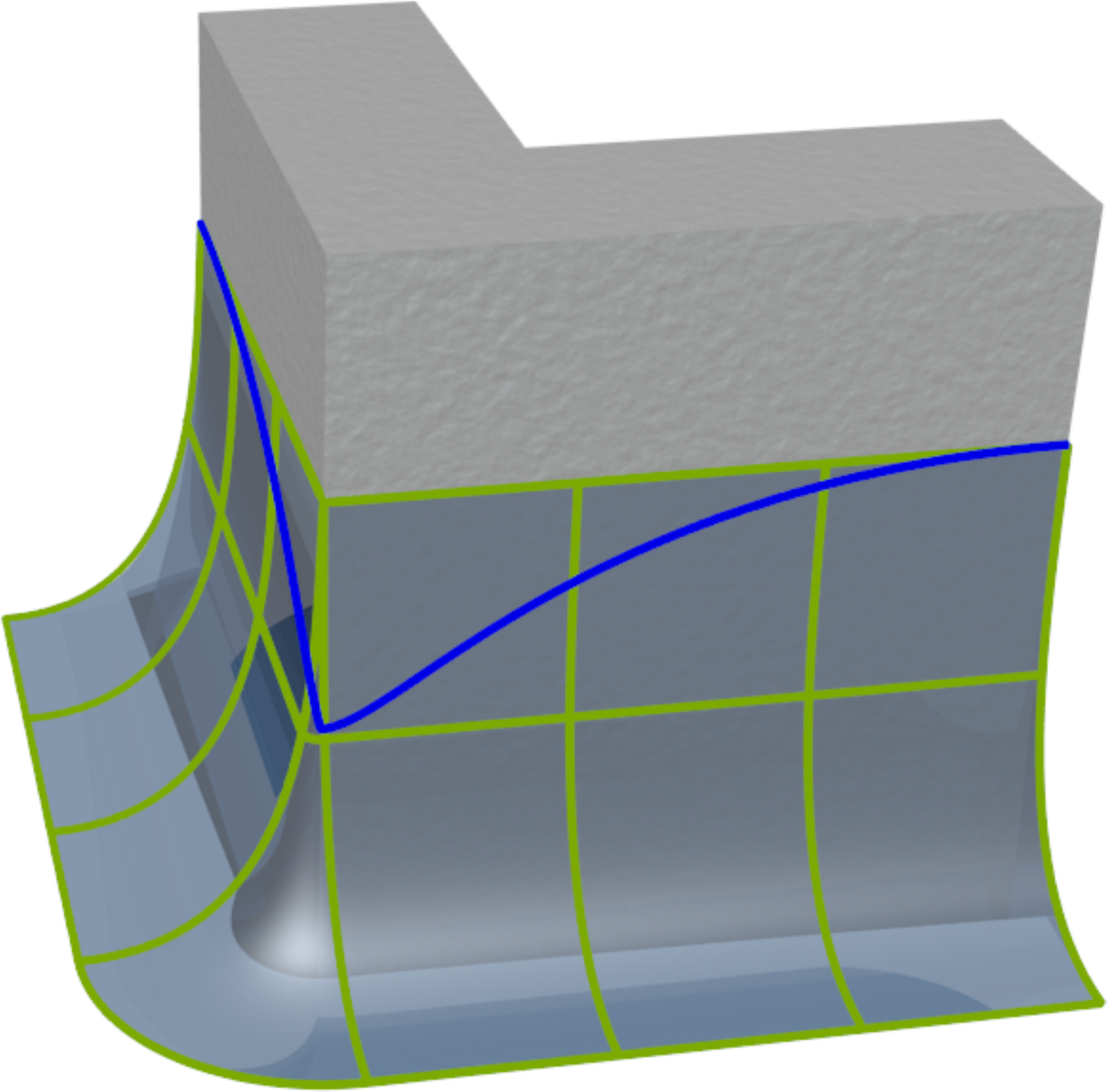}
\caption{A sketch showing the  situation of two menisci meeting at a corner without a cusp (green lines) and the expected contact line contour of a cusp (blue line).}
\label{fgr:cuspExplanation}
\end{figure}

Figure \ref{fgr:cuspExplanation} shows in more detail the contact line contour at an outer corner. If the corner would have no affect on the contact line at the point where the contact lines of both faces meet, the radius of curvature (green lines) would become zero, which would  induce an infinite capillary pressure into the liquid. This is physically impossible and leads to the formation of a cusp (blue line), creating a two-dimensional curvature of the liquid surface, balancing the  capillary pressure and the gravitational force.

\subsection{Interaction of the two corners}

\subsubsection{Contact line contours}

\begin{figure}
\centering
\includegraphics[width=\textwidth]{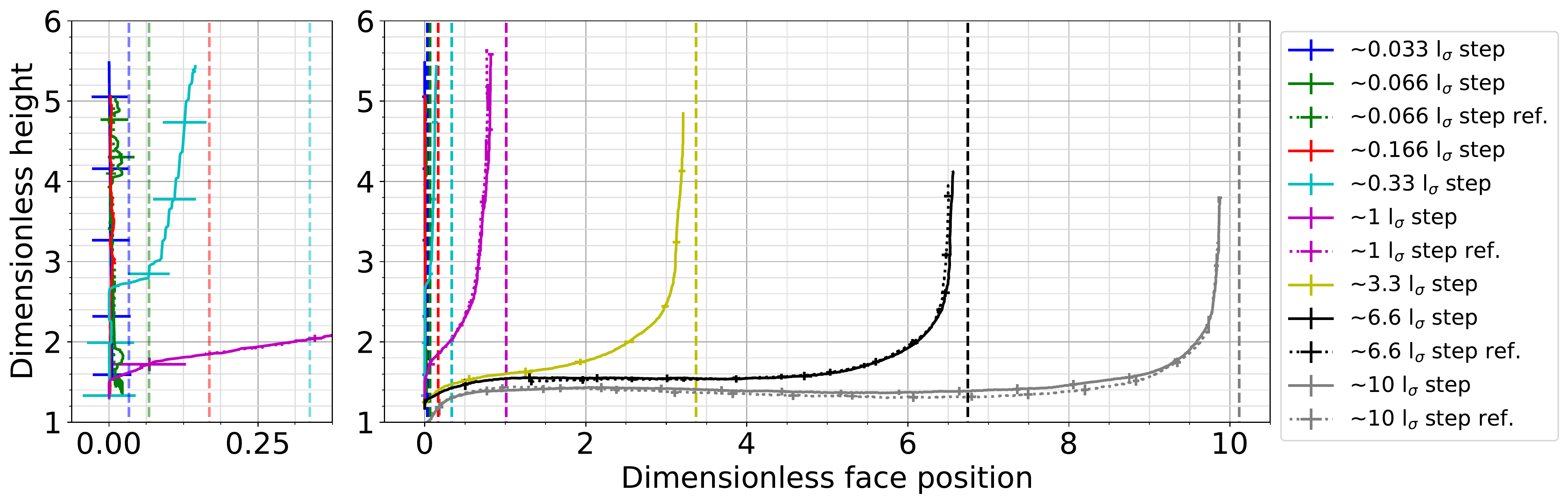}
\caption{The contact line on the face between the two corners for all sample sizes. The plots are aligned at the left side with the outer corner and all lengths are divided by the capillary length. The small plot on the left side shows the contact lines for the smallest step sizes but with an expanded x axis.}
\label{fgr:cuspFace}
\end{figure}

Figure \ref{fgr:cuspFace} shows the measured contact line contour on the step face between the outer corner on the left and the inner corner at the right. 
The right and left face of the samples are perpendicular to the plotted face and therefore not visible in this diagram. All lengths are divided by the capillary length to provide dimensionless data. The graphs are aligned at the outer corner (cusp at the left side), which is defined as face position zero. The vertical dashed lines show the inner corner positions for the different step sizes. The error bars represent one standard deviation computed from five measurements performed for every sample. The dotted lines show a second set of post processing results for comparison of systematic errors due to manual contact line detection. They are denoted \textit{ref.}  in the legend and show no systematic differences. For clarity, the smaller plot at the left side shows the contact line contour  for the five smallest step sizes, but with an expanded x axis.

For the largest analyzed step size ($\sim$10 l$_{\sigma}$), the cusp on the left and the rivulet on the right side appear to be unaffected from one another, although a very slight S-shape can be seen in-between, whose amplitude is of the order of one standard deviation of the measurement scatter. For all smaller step sizes the cusp and the rivulet show a clear interaction with an S-shaped curvature directly connecting the cusp with the rivulet. With decreasing step size the depth of the cusp also decreases. 

When the step size is $\sim$1 l$_{\sigma}$ or smaller, a new phenomenon appears, which is visible in the left smaller plot. While on the left sample face (compare Figure \ref{fgr:contactLine}) a cusp still appears, which is not visible in Figure \ref{fgr:cuspFace}, the contact line pins at the outer corner and follows it vertically up to a certain height, at which it separates from the edge and forms a rivulet without showing the S-shape seen for larger step sizes. While for $\sim$1 l$_{\sigma}$ the measurements exhibited different results switching between outer corner pinning and S-shaped contact lines, for $\sim$0.33 l$_{\sigma}$ the pinning at the outer corner becomes clearly evident in all measurements. If the step size is even smaller, the contact line stays pinned at the outer corner over the entire field of view, which can be observed for step sizes below $0.33$ l$_{\sigma}$. Due to the resolution limit of the camera and imperfections of the experimental samples, this contact line appears somewhat noisy in the stretched plot.

\subsubsection{Cusp behavior}

\begin{figure}
\centering
\includegraphics[width=0.75\textwidth]{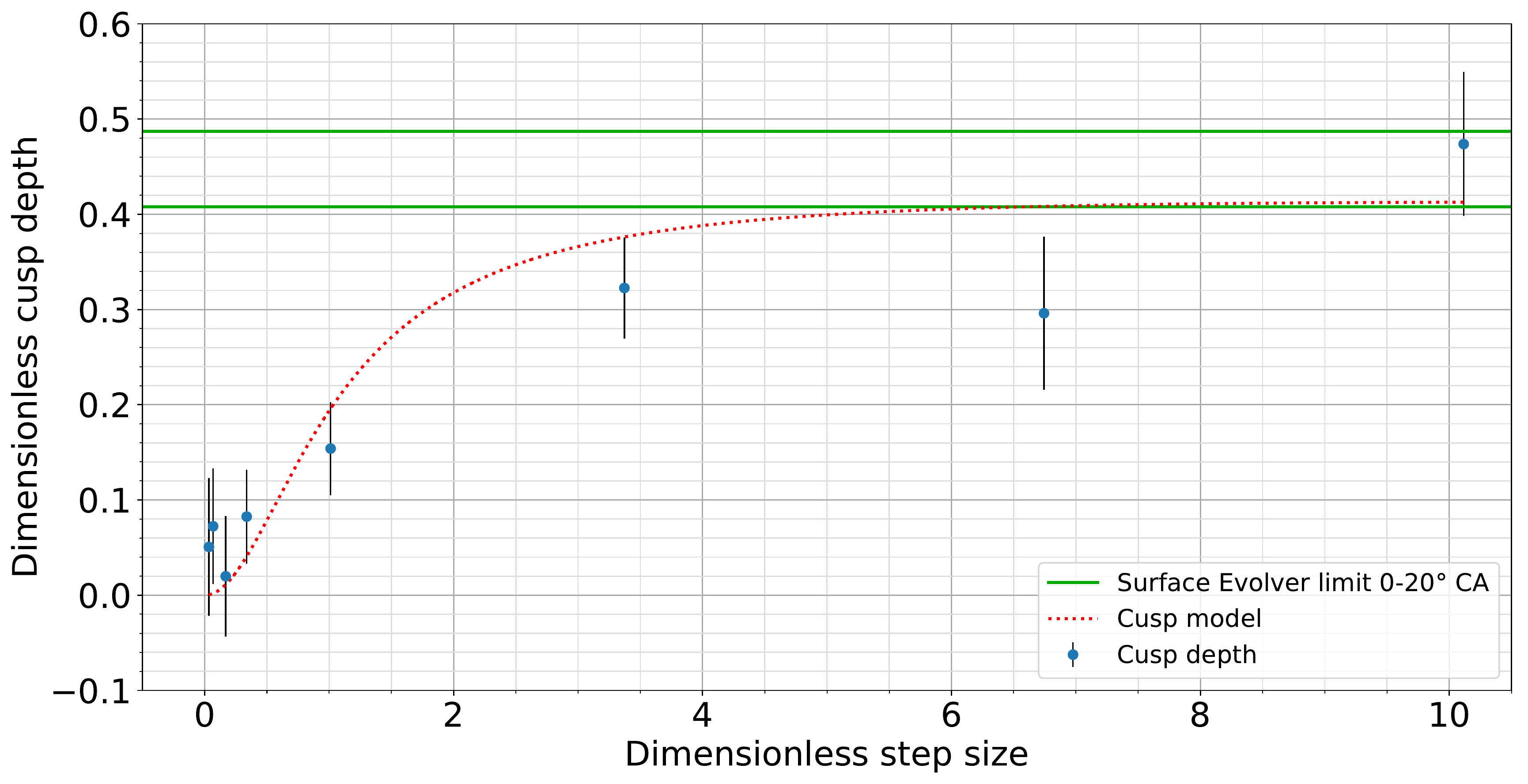}
\caption{The depth of the cusp below the average contact line height for the different step sizes. The green lines show the upper limit calculated with Surface Evolver for contact angles of $0^\circ$ and $20^\circ$. The dotted red line shows the cusp depth predicted by the model from equation \ref{eqn:cuspDepth} with the lower numerical limit chosen as $\text{c}_\text{max}$. All lengths are normalized with the capillary length.}
\label{fgr:cuspDepth}
\end{figure}

In Figure \ref{fgr:cuspDepth} the observed dimensionless cusp depth is plotted against the dimensionless step size. The cusp depth is defined as the vertical distance between the tip of the cusp and the average contact line height on the left and right faces.
It can be seen that for a vanishing step size the dimensionless cusp depth becomes zero.
This leads to a horizontal contact line at the left face of the sample, which becomes vertical and pinned on the outer corner, following it upwards until it separates from the corner to build the rivulet (compare to
$\sim$0.33 l$_{\sigma}$ from Figure \ref{fgr:cuspFace}). A similar result for groove geometries was obtained numerically by Thammanna Gurumurthy et al. \cite{ThammannaGurumurthy-2018-ID230}.
For an increasing step size the cusp depth increases.
From Figure \ref{fgr:cuspDepth} it can be assumed that the maximum cusp depth reaches an asymptotic value for an infinite step size.

To verify this, Surface Evolver, a software to calculate static liquid shapes by minimizing their free energies, is employed \cite{Brakke-1992-ID295}.
Note that because of its ability to solve static liquid surface shapes, Surface Evolver cannot be used to simulate the entire step geometry due to the presence of the infinite rising rivulet. Hence the computational domain consists of an outer corner with a wall length of 50 mm ($\sim$33.78 l$_{\sigma}$) in both directions in order to analyse the deepest possible cusp formation without any boundary effects.
In the direction perpendicular to the wall the liquid pool width is set to 30 mm ($\sim$22.27 l$_{\sigma}$).
To assure accurate calculations, the maximum final cell length for the mesh around the corner is set to 100 $\mu$m and the surface is evolved until the coefficient of variation of the last seven iteration steps is smaller than $10^{-7}$.
The contact angle at the wall is varied between $0^\circ$ and $20^\circ$ to reflect the uncertainty of the contact angle of silicon oil on the samples.
The calculation results in a maximum cusp depth of 0.487$\pm$0.0008 $l_{\sigma}$ and 0.408$\pm$0.0006 $l_{\sigma}$ for a contact angle of $0^\circ$ and $20^\circ$ respectively, which are plotted as horizontal green lines in Figure \ref{fgr:cuspDepth}.
This verifies the assumption that an asymptotic value for the cusp depth exists and that the maximum step size of $\sim$10 l$_{\sigma}$ is large enough to yield the limiting value of the deepest cusp.

To predict the cusp depth for different step sizes it is assumed that a cusp having the asymptotic depth limit is pulled upwards by the increased slope of the contact line due to the  rivulet. For comparison the slope of a rivulet unaffected by a cusp is used. Using this assumption and the slope of the unified rivulet (Eq.~(\ref{eqn:unified}) presented below), the depth of the cusp can be expressed as 

\begin{equation}
c_\text{depth} = \frac{\text{c}_\text{max}}{1-\frac{\partial h}{\partial x}}\label{eqn:cuspDepth}
\end{equation}

This equation is plotted in Figure \ref{fgr:cuspDepth}, using the 0.408 $l_{\sigma}$ limit from Surface Evolver as $\text{c}_\text{max}$. It can be seen that the model fits the measured cusp depth  well, except for the second step size from the right and for the second step size from the left, which both exhibit some deviation.  Since there exists no physical reason for single step sizes to show such deviations, it is assumed that this discrepancy results from inaccuracies in the calibration.

\subsubsection{Rivulet behavior}

\begin{figure}
\centering
\includegraphics[width=\textwidth]{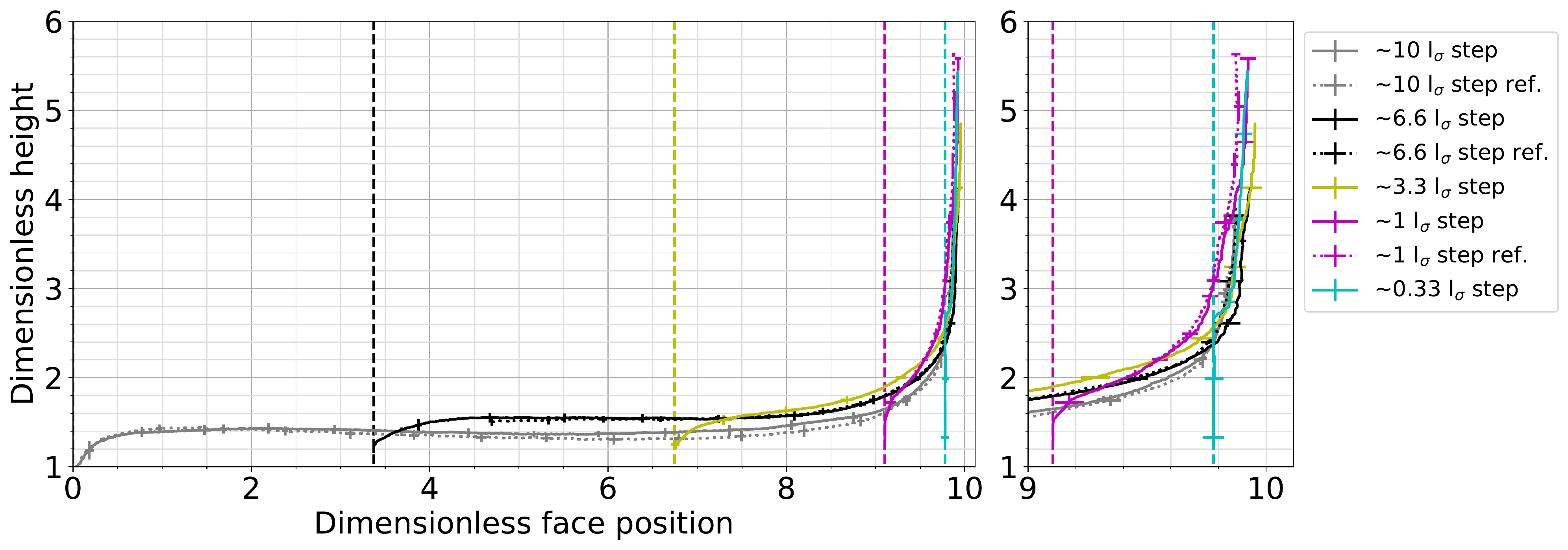}
\caption{The contact line on the face between the two corners. Only samples larger than $\sim0.166$ l$_{\sigma}$ are shown, since smaller step sizes lead to complete pinning at the outer corner and distortion of the rivulet shape (compare Figure \ref{fgr:cuspFace} on the left). The plots are aligned at the inner corner, represented by the right end of the plot at a dimensionless face position of 10. All lengths are normalized by the capillary length.}
\label{fgr:rivuletFace}
\end{figure}

Figure \ref{fgr:rivuletFace} shows the same data as Figure \ref{fgr:cuspFace}, but without showing the completely pinned contact lines on the outer corner for very small step sizes and with the face position aligned with the inner corner at the value 10 (right side of the graph). The respective outer corners are visualized as vertical dashed lines. With this data representation it becomes obvious that all rivulets exhibit the same shape, independent of the step size. This is more or less valid even for the $\sim$0.33 l$_{\sigma}$ step, which is vertically pinned on the outer corner until it crosses the static rivulet shape of the larger steps. At this point the contact line detaches from the outer corner and follows the same rivulet shape as all other step sizes. Hence, the general shape of the rivulet close to the inner corner is independent of the step size and should be describable by a universal function.

Ponomarenko et al.\cite{Ponomarenko-2011-ID202} described the rivulet shape by using the capillary tube model from Figure \ref{fgr:rivuletExplanation}a). The assumption of this theory is the fitting of round capillary tubes of radius $r$ between the walls of the corner and calculating the capillary rise for these tubes, given by the balance of the Laplace pressure for the corresponding {\it vertical} curvature with gravity. 

\begin{equation}
h = \frac{2\sigma cos(\theta)}{\rho \text{g} r} \label{eqn:tubeRise}
\end{equation}

\begin{figure}
\centering
\includegraphics[width=0.75\textwidth]{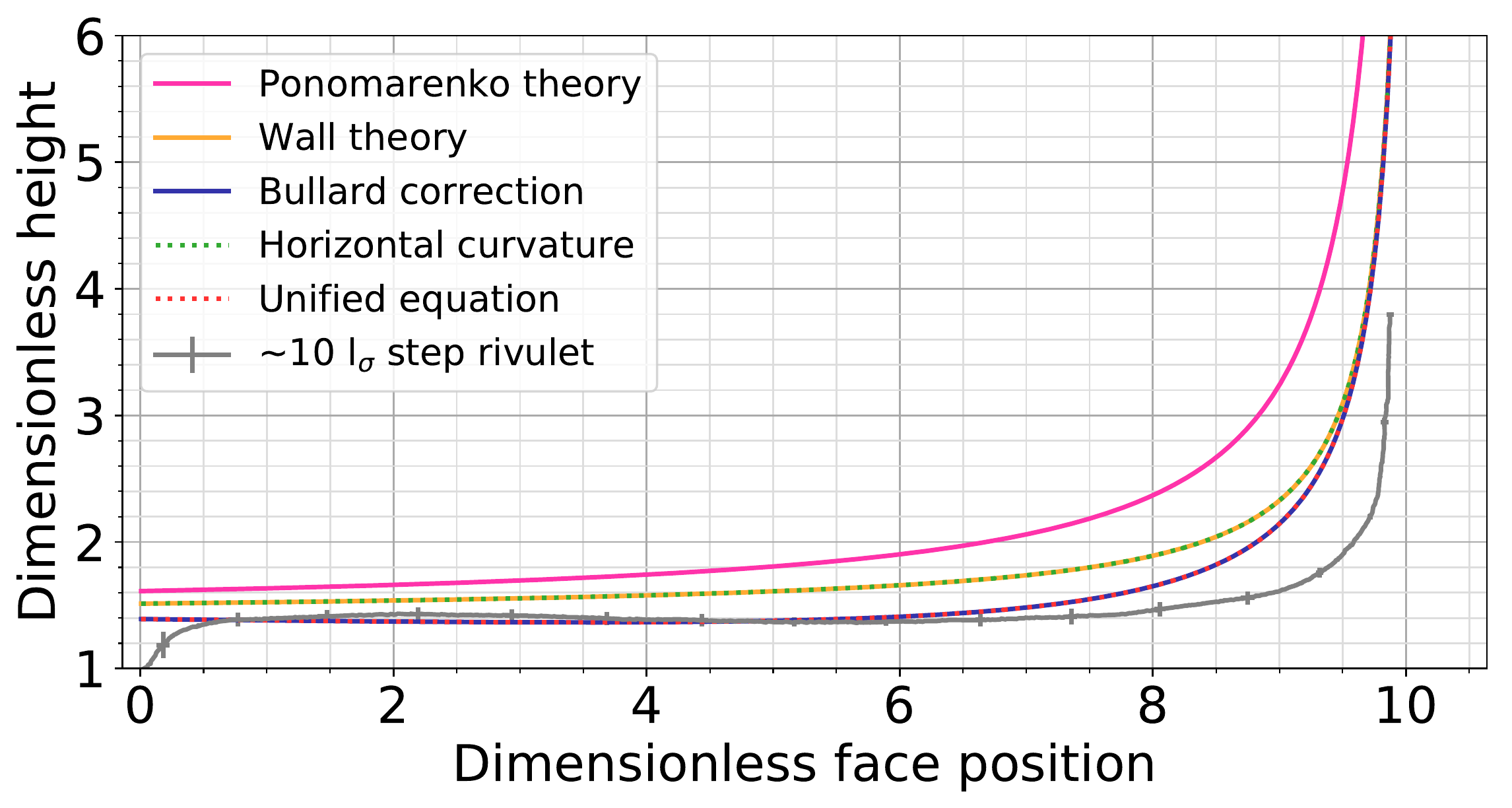}
\caption{The contact line of the $\sim$10 l$_{\sigma}$ step and different relations for its shape with a contact angle of 0$^{\circ}$.  The static meniscus rise height is added as an offset to all results (seen at face position zero).}
\label{fgr:rivuletEqs}
\end{figure}

 \noindent  This rise height in the tubes was then used as the rivulet rise height at the wall positions contacted by the virtual tubes. The resulting rivulet shape  is compared to the measurement data in Figure \ref{fgr:rivuletEqs}. The tube radius $r$  for the studied 270$^{\circ}$ inner corner is set to the distance from the corner along the wall to the contact line position  along the horizontal.

Bullard et al. \cite{Bullard-2009-ID289} performed numerical simulations of the meniscus shape between parallel walls and reported the position of the deepest meniscus point relative to the average height calculated with the equation 

\begin{equation}
h = \frac{2\sigma cos(\theta)}{\rho \text{g} w} \label{eqn:wallRise}
\end{equation}

\noindent This equation assumes parallel walls instead of circular tubes and can be used in the same way as Eq.~\ref{eqn:tubeRise} to describe the rivulet shape, which was already done by O'Brien et al. \cite{O'Brien-1968-ID293}. In equation \ref{eqn:wallRise} the  perpendicular distance between the walls $w$ ($=2x$)  is used, leading to different rivulet shapes in the inner corner if Eqns.~\ref{eqn:tubeRise} and \ref{eqn:wallRise} are used for the same corner geometry. To compare results arising from the tube and wall rise expressions, Eq.~\ref{eqn:tubeRise} is plotted as "Ponomarenko theory" and Eq.~\ref{eqn:wallRise} is plotted as "Wall theory" in Figure \ref{fgr:rivuletEqs}. Since the average $h$ obtained from these analytical equations is only known to be between the deepest meniscus point and the contact line and its relative position  changes with the surface shape, it contains an inherent error in the length scale of the meniscus height. To correct this error Bullard et al. fitted an empirical equation to their numerical results in order to obtain an equation for the deepest point of the meniscus between two parallel walls. The result is Eq.~\ref{eqn:bullardEq}, plotted in Figure \ref{fgr:rivuletEqs} as "Bullard correction". It is obvious that the correction applied by this equation is critical for the wall distances analysed in the current study, since the contact line height differs significantly from the height predicted by Eq.~\ref{eqn:wallRise}.

\begin{equation}
\begin{split}
h &= \left(\frac{2 cos(\theta)}{\sigma_{\scalebox{.5}{B}}}-f(\theta)g(\sigma_{\scalebox{.5}{B}})e^{-4.48\sigma_{\scalebox{.5}{B}}^{\frac{1}{8}}}\right) w \\
\sigma_{\scalebox{.5}{B}} &= \frac{\rho g w^2}{\sigma} \\
f(\theta) &= 9.24cos(\theta)+2.13cos(\theta)^3 \\
g(\sigma_{\scalebox{.5}{B}}) &= 0.834\sqrt{\sigma_{\scalebox{.5}{B}}}-0.024\sigma_{\scalebox{.5}{B}}
\label{eqn:bullardEq}
\end{split}
\end{equation}

The model from Fig.~\ref{fgr:rivuletExplanation}b) is plotted in Fig.~\ref{fgr:rivuletEqs} as "Horizontal curvature" and is given by the equation

\begin{equation}
h = \frac{\sqrt{2} \sigma sin\left( \frac{\pi}{4}-\theta\right)}{\rho \text{g} x} \label{eqn:pressEq}
\end{equation}

\noindent In this  equation  the horizontal distance from the corner to the contact line, $x$, is used to describe the radius of curvature in the {\it horizontal} direction. Although the curves for Eqns.~\ref{eqn:wallRise} and  \ref{eqn:pressEq} are identical for 0$^{\circ}$ contact angle they differ for higher contact angles, as mentioned in the subsequent section ``General phenomena at the two corners''. Equation \ref{eqn:wallRise} results in an infinite rivulet for all contact angles below 90$^{\circ}$, while Eq.~\ref{eqn:pressEq} predicts an infinite rivulet for contact angles below 45$^{\circ}$, a flat liquid interface for 45$^{\circ}$ and a negative infinite rivulet for higher contact angles.

Both models from Fig.~\ref{fgr:rivuletExplanation}  represent only one half of the real physical problem. The capillary tube or alternatively parallel wall rise model from Fig. \ref{fgr:rivuletExplanation}a) accounts only for the vertical curvature of the liquid. Close to the corner where the steepness of the rivulet becomes very strong, the vertical curvature of the rivulet becomes very low, while the only remaining curvature is in the horizontal direction. The model from Figure \ref{fgr:rivuletExplanation}b) accounts only for this horizontal curvature, but neglects the vertical curvature further away from the corner.

To distinguish between the horizontal and the vertical curvature of the rivulet, the slope of the rivulet with respect to the horizontal plane can be assessed. It is described as $\alpha$ in 

\begin{equation}
\alpha = arctan\left(-\dfrac{\partial h}{\partial x}\right) \label{eqn:alpha}
\end{equation}

\noindent which uses a negative sign in order to result in positive angles for the increasing rivulet height $h$ at decreasing corner distances $x$.
A simple combination of Eqns.~\ref{eqn:tubeRise}, \ref{eqn:wallRise} or \ref{eqn:bullardEq} with Eq.~ \ref{eqn:pressEq} using $cos(\alpha)$ and $sin(\alpha)$, or even $cos(\alpha)^2$ and $sin(\alpha)^2$ to preserve a weighting sum of 1, is not appropriate, since  numerical analyses of the resulting differential equations gave rise to highly irregular rivulet shapes, which appear to result from several or infinite valid solutions of the differential equations. The same effect appears when the equations are added to each other without weighting, and for the contact angle   the apparent contact angle from the front or respectively the top is used, depending on $\alpha$.

\begin{figure}
\centering
\includegraphics[width=0.3375\textwidth]{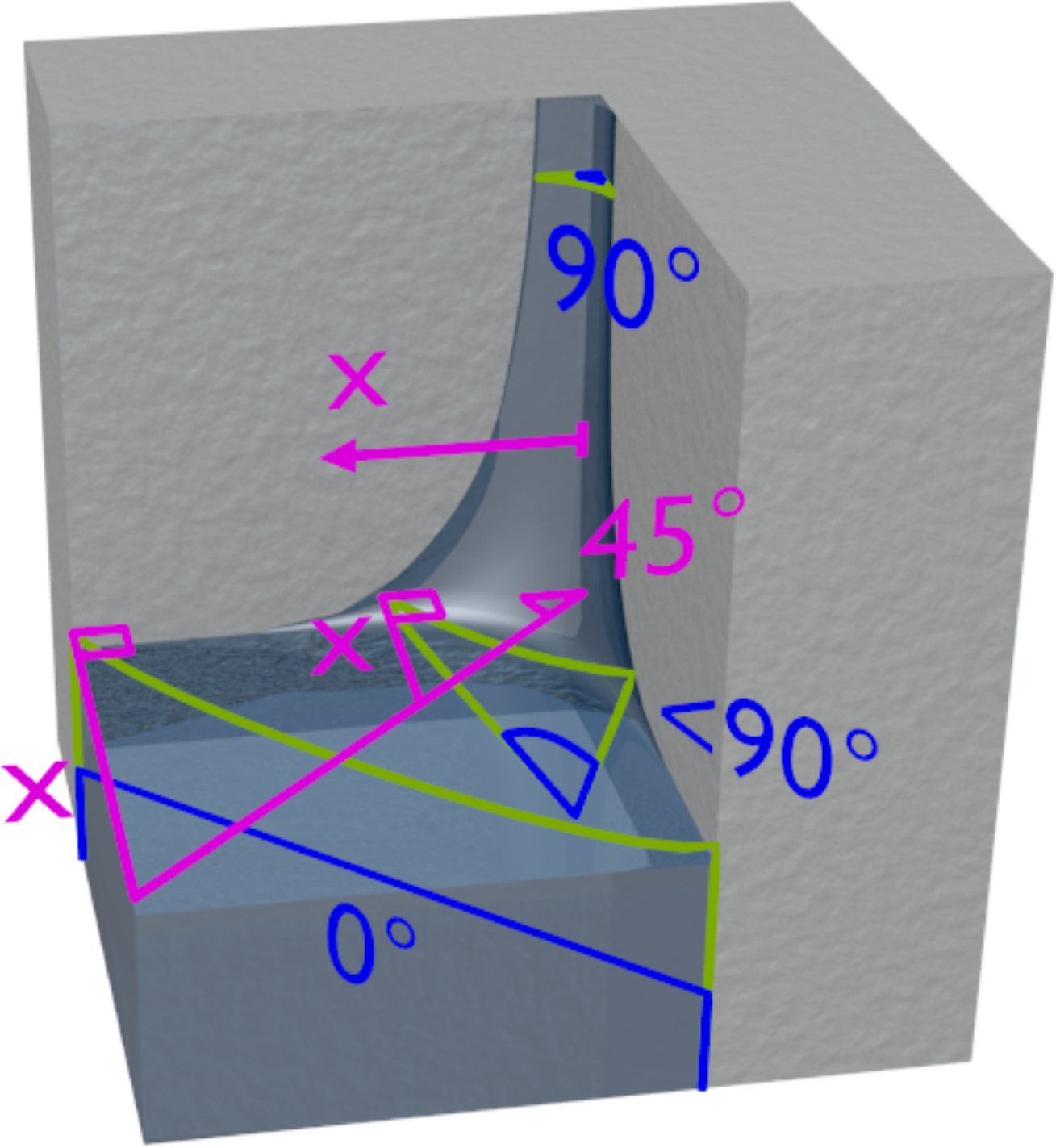}
\caption{Three liquid slides in a  direction normal to the liquid surface are shown for a rivulet. The wall opening angle for each slice is shown in blue. The angle bisector of the corner and connection lines perpendicular to the wall, having always the same length as their corresponding wall position $x$, are shown in magenta.}
\label{fgr:uniEqExp}
\end{figure}

Therefore an equation has been developed, combining both model assumptions of Fig. \ref{fgr:rivuletExplanation} into a single differential equation with only one valid solution. The first constraint for deducing an unified equation is the fact that for a 270$^{\circ}$ inner corner the perpendicular distance from the current wall position $x$ to the angle bisector of the corner (45$^{\circ}$ plane) is always equal to the wall position itself, as visualized in magenta in Figure \ref{fgr:uniEqExp}. The second constraint  concerns the calculation of the rise height for individual liquid slices in the  direction normal to the rivulet. Figure \ref{fgr:uniEqExp} shows three such slices in green. These slices point in the direction of the main curvature of the liquid. Tests including the second curvature along the contact line into the rivulet calculation showed that its effect on the rivulet shape is negligible, supporting the model of main curvature liquid slices. 

Each slice is positioned between two walls of distance $x$, as known from the first constraint. Due to their inclination they experience different angles between the two walls of the corner as shown in blue in Figure \ref{fgr:uniEqExp}. The vertical slices perceive two infinite parallel walls with an opening angle of 0$^{\circ}$ in-between. This opening angle increases with increasing inclination, making the walls meet. When the slices become horizontal, the opening angle reaches 90$^{\circ}$. The angle between the two walls is equal to $\alpha$ for all slices. Calculating the rivulet rise height using these constraints results in the equation

\begin{equation}
h = \frac{\sigma sin\left(\frac{\pi}{2}-\theta-\frac{\alpha}{2}\right)}{cos\left(\frac{\alpha}{2}\right) \rho \text{g} x} \label{eqn:unified}
\end{equation}

\noindent For $\alpha = 0^{\circ}$, far away from the corner, equation \ref{eqn:unified} becomes equation \ref{eqn:wallRise}. When $\alpha = 90^{\circ}$, close to the corner, the equation becomes equal to Eq.~\ref{eqn:pressEq}. Equation \ref{eqn:unified} is a first-order nonlinear ordinary differential equation (when substituting Eq.\ref{eqn:alpha} for $\alpha$  with no analytical solution known to the authors. Hence, it has to be solved numerically for which the equation has shown a stable behavior.

Like Eq.~\ref{eqn:wallRise}, Eq.~\ref{eqn:unified} describes an average $h$, whose  relative position inside the liquid meniscus changes with the surface shape. To make use of the empirical height correction of Bullard et al., Eq.~\ref{eqn:bullardEq} has been compared to Eq.~\ref{eqn:wallRise}. It can be seen that Eq.~\ref{eqn:bullardEq} is identical to Eq.~\ref{eqn:wallRise}, except for an additional empirical correction term. This empirical correction term was added to Eq.~\ref{eqn:unified} in order to obtain 

\begin{equation}
h = \frac{\sigma sin\left(\frac{\pi}{2}-\theta-\frac{\alpha}{2}\right)}{cos\left(\frac{\alpha}{2}\right) \rho \text{g} x}-2x\left(f(\theta)g(\sigma_{\scalebox{.5}{B}})e^{-4.48\sigma_{\scalebox{.5}{B}}^{\frac{1}{8}}}\right) \label{eqn:unifiedBullard}
\end{equation}

\noindent plotted as "Unified equation" in Fig.~\ref{fgr:rivuletEqs}. Technically speaking the correction should only be applied to the part of the rivulet following the model of Fig.~\ref{fgr:rivuletExplanation}a). Due to the vanishing offset from the correction term close to the corner, where the model of Fig.~\ref{fgr:rivuletExplanation}b) becomes dominant, a further blending of the correction term is omitted in this equation.

\begin{figure}
\centering
\includegraphics[width=0.75\textwidth]{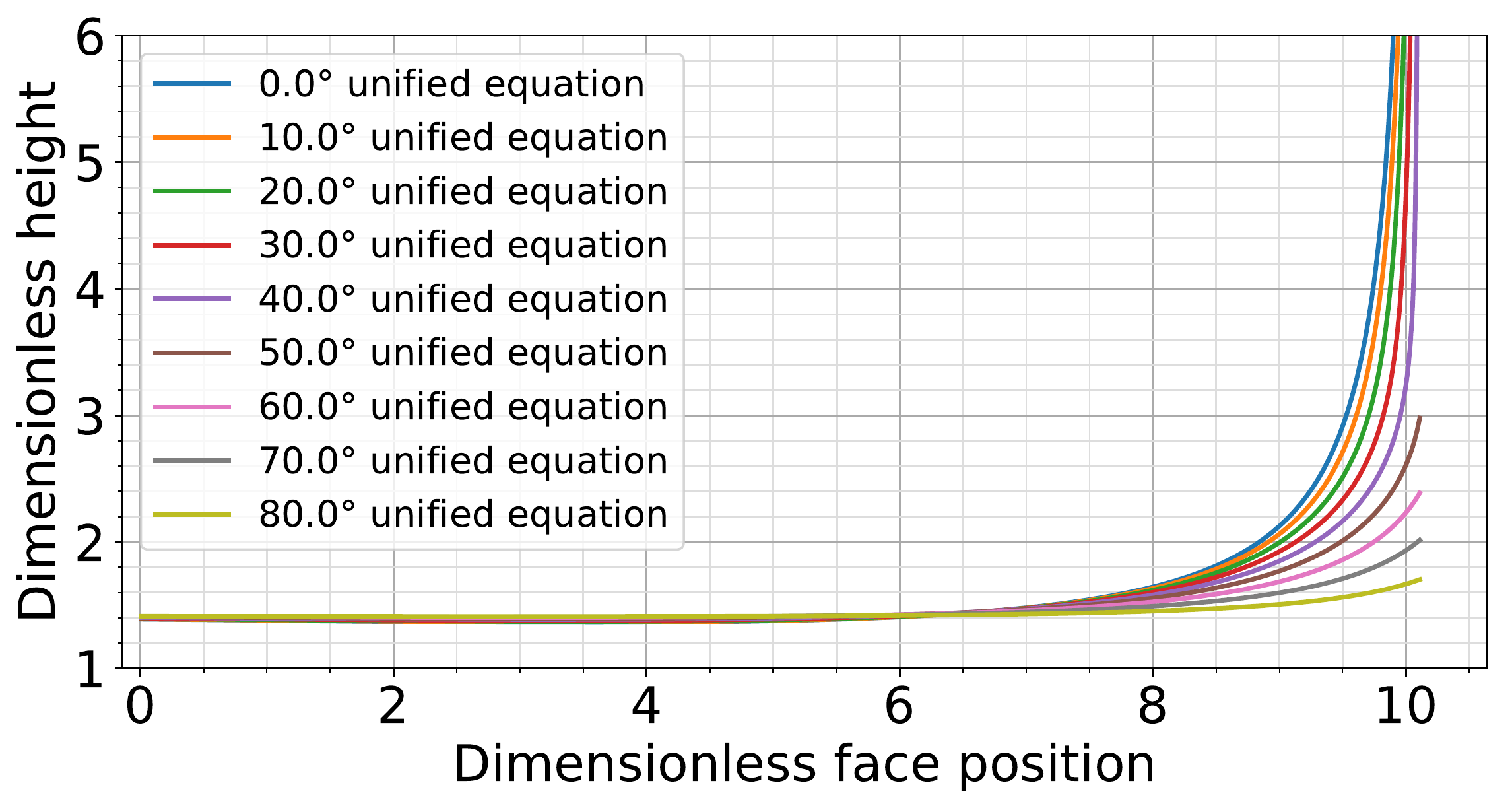}
\caption{The results of Eq.~\ref{eqn:unifiedBullard} for different contact angles with a fixed offset value for comparison.}
\label{fgr:unifiedEquationPlot}
\end{figure}

Eqns.~\ref{eqn:bullardEq} and  \ref{eqn:unifiedBullard} appear as a single curve in Fig.~\ref{fgr:rivuletEqs}, the difference between the two equations becomes visible for different contact angles. While Eq.~\ref{eqn:bullardEq}, like the uncorrected Eq.~\ref{eqn:wallRise}, predicts an infinite rivulet rise for every contact angle below 90$^{\circ}$, Eq.~\ref{eqn:unifiedBullard} is the first rivulet shape description known to the authors which fulfills the Concus-Finn criteria \cite{Concus-1969-ID294}. It correctly predicts an infinite rivulet rise for contact angles below 45$^{\circ}$ and results in finite rivulet rise heights for contact angles above 45$^{\circ}$, approaching a flat liquid surface at 90$^{\circ}$ contact angle. Figure \ref{fgr:unifiedEquationPlot} shows the results of Eq.~\ref{eqn:unifiedBullard} for different contact angles. While the finite rivulet rise heights for contact angles above 45$^{\circ}$ qualitatively fit to the rivulet rise heights observed in preliminary experiments, the authors could not achieve reproducible contact angles above 45$^{\circ}$ to quantitatively compare the real world rise heights with the theoretical values from Eq.~\ref{eqn:unifiedBullard}.

\subsection{Rivulet dynamics}

The rise of the rivulet tips is tracked during the experiments in order to calculate the rivulet rise velocicty. While all lengths in the current study are made dimensionless by dividing them by the capillary length l$_{\sigma}$, the times are non-dimensionalized by dividing them by a characteristic time t$_{\sigma}$ which is defined as $\tfrac{\mu}{\sqrt{\rho \text{g} \sigma}}$, in which $\mu$ is the dynamic viscosity of the  liquid. Dividing the distance per time step by l$_{\sigma}$ and the time step by t$_{\sigma}$ has the same effect as multiplying the measured rivulet velocity with $\tfrac{\mu}{\sigma}$ in order to receive a dimensionless velocity.

\begin{figure}
\centering
\includegraphics[width=0.75\textwidth]{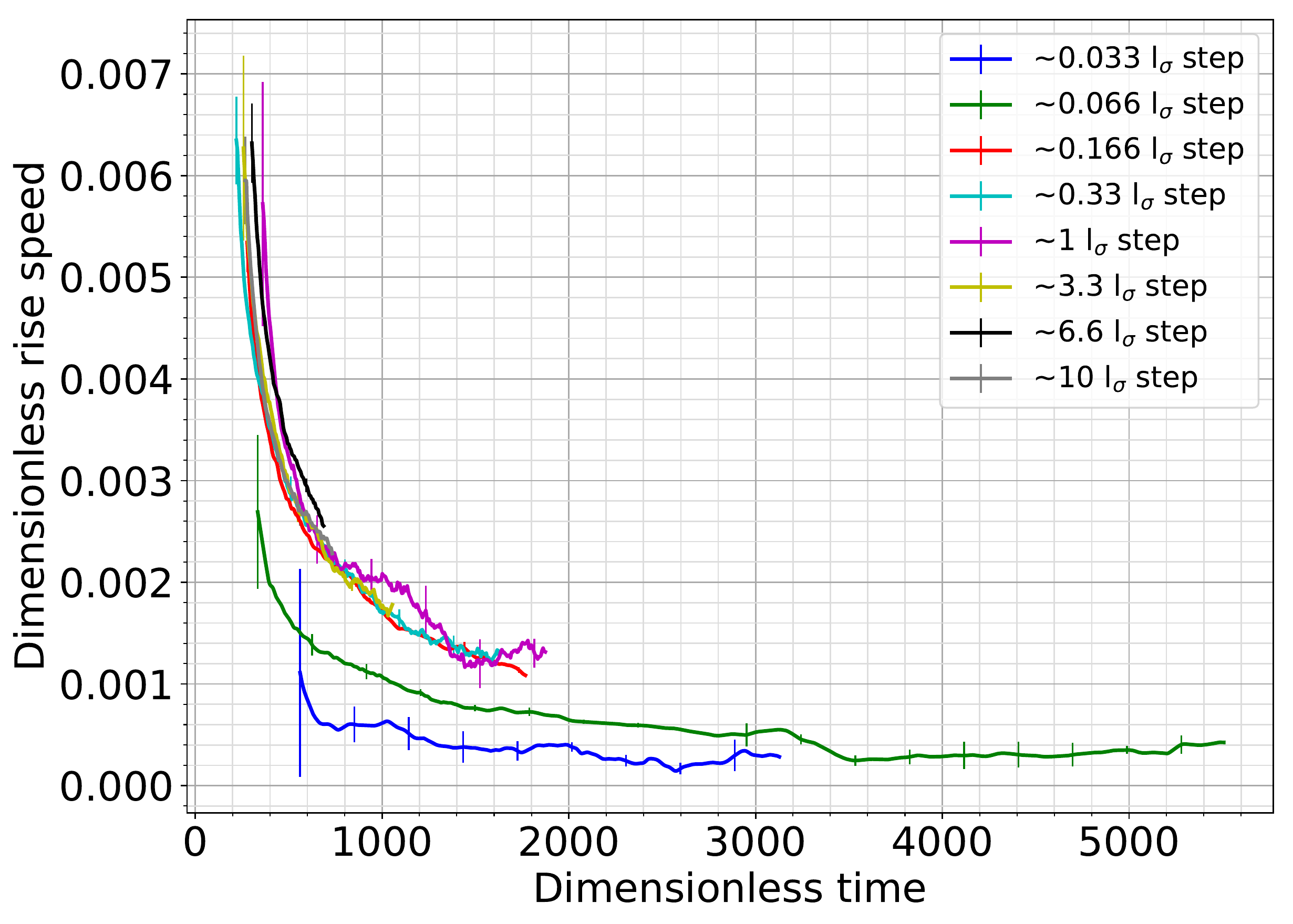}
\caption{The dimensionless velocity of the rivulet tip.}
\label{fgr:rivuletSpeed}
\end{figure}

The measured raw data contains noise which results in fluctuating speeds; hence, the data is post-processed to reduce these fluctuations. In a first step the raw data is interpolated using cubic splines. Afterwards the derivatives of these splines are calculated and evenly sampled with fixed time steps of twice the frame rate of the original camera video (249,986 fps). In a following step a moving average is applied to the resampled data, using a window width of a quarter of the original frame rate to average steps of half a second of duration. Data closer than half of the averaging window to the borders of the data sets were discarded. Figure \ref{fgr:rivuletSpeed} shows the resulting velocity plots displaying the decreasing rivulet tip velocity over time.

The rivulet tip rise velocity is, except for the smallest two step sizes, independent of the step size. This is an unexpected result since Figure \ref{fgr:rivuletFace} shows that the rivulet is truncated close to the corner for step sizes $\sim$ $\le$ l$_{\sigma}$, leading to  asymmetric  rivulet shapes. When comparing these boundary conditions to the model from Figure \ref{fgr:rivuletExplanation}b), the horizontal liquid slices have to change their curvature in order to exhibit the contact angle on one face and the pinning corner position on the other face. Since the curvature is the only driving force of the rivulet, the observed  rising behavior suggests that only the rivulet region close to the tip is relevant for the rivulet rise, while the influence of the rest of the rivulet can be neglected.  Furthermore, the rivulet tip never stops rising, as predicted by Ponomarenko et al.\cite{Ponomarenko-2011-ID202}; hence, it always provides a driving force upwards. 

At small step size the rivulet rise velocity is lower and also exhibits higher fluctuations. The increased fluctuations in the speed  do not only arise from  measurement noise, but also from increased interaction of the rivulet with surface roughness, which becomes capable of locally slowing down the rivulet  and speeding it up again. However, to explain the lower rise velocity at very small step size some change in the rivulet tip itself must be postulated. 

A very simple model for this observation was developed in which  the region close to the rivulet tip was assumed to have a similar size as the final rivulet width. Since the variation  in rivulet width becomes small on top of its base (compare Figure \ref{fgr:rivuletFace}), the rivulet tip size was assumed to be nearly constant in the observed sample region of less than six l$_{\sigma}$ rise heights. Since the rivulet tip at the start of the rise is smaller than the final rivulet base width, it is assumed that the rivulet tip size is constant from the beginning of the rise. To further simplify the model, only one characteristic horizontal rivulet slice is considered to be representative for the force balance in the rivulet tip. The friction forces are assumed to be of comparable size for an unaffected and an asymmetrically pinned rivulet. Gravitational forces are negligible due to the width of the rivulet tip being $<$ l$_{\sigma}$, which means the ratio of gravitational forces to surfaces forces is smaller than one. This results in a model only comparing the mean curvature of an unaffected and a pinned liquid slice to estimate the decrease in rivulet rise speed. Both rivulet slices are assumed to be perfect arcs.

Due to the small size and thickness of rivulets on very small steps, the rivulet width during rise cannot be observed in a reliable way and therefore must be considered unknown. From this point on there are two different assumptions which can be made. Either the width of the rivulet tip on the right wall can be assumed to be constant, called $d$, or, since the model neglects changes in the friction force which are caused by wall interaction, the overall wetted wall length of the rivulet tip, which is $2d$, can be assumed to be constant. Accordingly, for the latter case the contact line position at the right wall changes to $d^* = 2d-s$ for a step size of $s$ smaller than $d$. At the right wall the rivulet tip has to exhibit its original contact angle. At the same time the rivulet arc should meet the point of the outer corner were it is pinned. Due to the pinning, the contact angle at this point is not defined. With this constraint a rivulet arc can be constructed, having, depending on the right wall width assumption, a radius described by the equation 

\begin{equation}
\begin{split}
d = \text{const:} \quad r &= -\frac{d^2+s^2}{2(sin(\theta) d - cos(\theta) s)} \qquad \text{for} \quad s \leq d \quad \text{and} \quad \theta < 90^{\circ} \\
2d = \text{const:} \quad r &= -\frac{2d^2-2ds+s^2}{sin(\theta) d - cos(\theta) s} \qquad \text{for} \quad s \leq d \quad \text{and} \quad \theta < 90^{\circ} \label{eqn:r}
\end{split}
\end{equation}

For a symmetric arc without a pinning edge $s$ equals $d$. The curvature of the surface $\kappa$ is inverse to $r$, resulting in Eq.~\ref{eqn:speedCorr} as a scaling factor to compare the rivulet rise speeds of very small and larger step sizes.

\begin{equation}
\begin{split}
d = \text{const:} \quad \frac{\kappa_{\text{unaffected}}}{\kappa_{\text{pinned}}} &= \frac{(d^2+s^2)(sin(\theta)-cos(\theta))}{2d(sin(\theta) d - cos(\theta) s)} \qquad \text{for} \quad s \leq d \quad \text{and} \quad \theta < 90^{\circ} \\
2d = \text{const:} \quad \frac{\kappa_{\text{unaffected}}}{\kappa_{\text{pinned}}} &= \frac{(2d^2-2ds+s^2)(sin(\theta)-cos(\theta))}{d(sin(\theta) d - cos(\theta) s)} \qquad \text{for} \quad s \leq d \quad \text{and} \quad \theta < 90^{\circ} \label{eqn:speedCorr}
\end{split}
\end{equation}

\begin{figure}
\centering
\includegraphics[width=\textwidth]{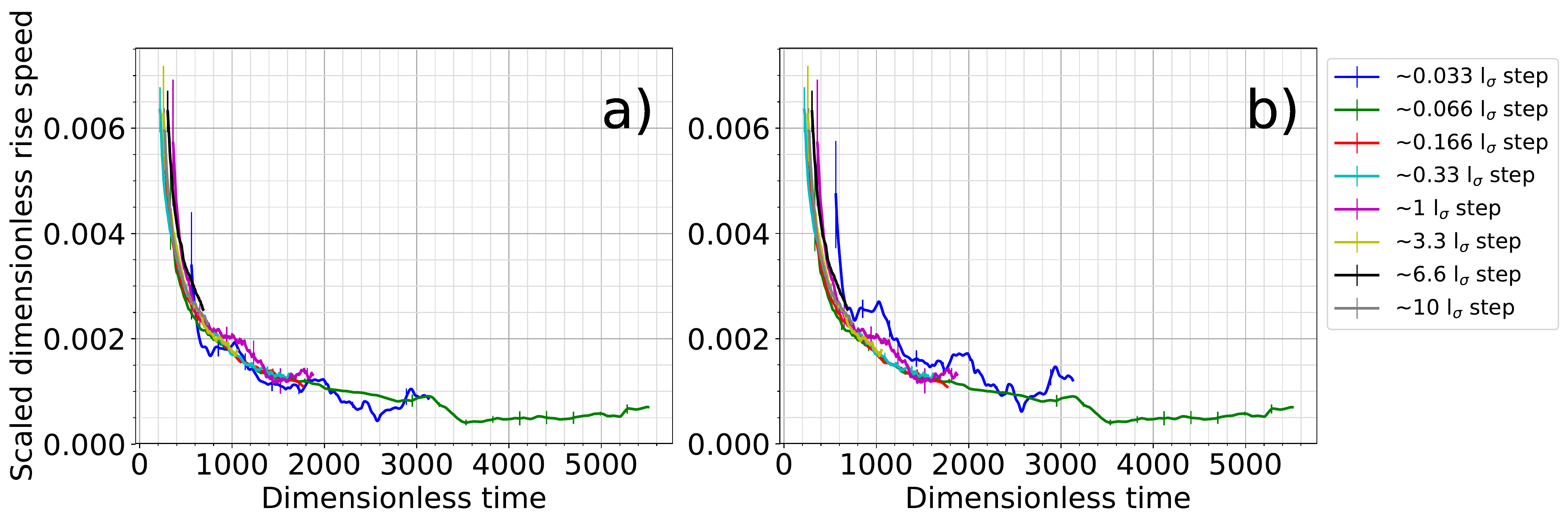}
\caption{The dimensionless velocity of the rivulet rise with the new model applied. a) Uses the assumption of a constant right wall width $d$ of 0.2 l$_{\sigma}$, b) uses the assumption of a constant overall wetting length, with a $d$ value of 0.1 l$_{\sigma}$.}
\label{fgr:rivuletSpeedCorrected}
\end{figure}

When $s < d$ and $\theta < 90^{\circ}$ the ratios given in equation \ref{eqn:speedCorr} are always $>$ 1. Multiplied by this factor the slower rise speeds should equal the unaffected rise speeds. Figure \ref{fgr:rivuletSpeedCorrected} shows the rivulet speed data with $d$ chosen to be 0.1 l$_{\sigma}$ for one and 0.2 l$_{\sigma}$ for the other version of the model. Both lengths of $d$ are in the range which can be estimated when observing Figure \ref{fgr:rivuletFace}. Although the developed model is very simple and it cannot be determined which of the two assumptions for the right wall is correct, it is capable of correcting the rivulet rise speeds in both versions using realistic values for $d$.

\section{Conclusions}

The current work observes the spontaneous wetting of samples comprising  three faces separated from each other with 270$^{\circ}$ inner and 90$^{\circ}$ outer corners.  It is shown that the rivulet is pinned at the outer corner when the distance between the outer and the inner corner becomes smaller than the capillary length. Furthermore, the cusp depth on the outer corner and its disappearance as the step size decreases towards zero is quantified. An empirical equation is proposed to describe the general relationship between cusp depth and step size. 
By comparing the results of two common models to describe the rise of rivulets, a new, unified physical model is deduced. The model described in equation \ref{eqn:unifiedBullard} is the first model able to quantitatively predict rivulet shapes for all contact angles between 0$^{\circ}$ and 90$^{\circ}$. Furthermore it is shown that the rivulet rise occurs rather independent of the step size. Only very small step sizes, being small enough to deform the leading rivulet tip, are able to influence the rivulet rise speed. For this phenomena a very simple theoretical model in two different versions is developed. Even though this model is of empirical nature, it is able to quantitatively describe the effect of small step sizes slowing down rivulet rise. With this observations and set of models for the different phenomena, the current work supports the understanding of rivulet and cusp formation and their interaction in close proximity to one another.

\begin{acknowledgement}

We kindly acknowledge the financial support by the German Research Foundation (DFG) within the Collaborative Research Centre 1194 ``Interaction of Transport and Wetting Processes'', Project A01.

\end{acknowledgement}


\bibliography{references}

\providecommand{\latin}[1]{#1}
\makeatletter
\providecommand{\doi}
  {\begingroup\let\do\@makeother\dospecials
  \catcode`\{=1 \catcode`\}=2\doi@aux}
\providecommand{\doi@aux}[1]{\endgroup\texttt{#1}}
\makeatother
\providecommand*\mcitethebibliography{\thebibliography}
\csname @ifundefined\endcsname{endmcitethebibliography}
  {\let\endmcitethebibliography\endthebibliography}{}
\begin{mcitethebibliography}{32}
\providecommand*\natexlab[1]{#1}
\providecommand*\mciteSetBstSublistMode[1]{}
\providecommand*\mciteSetBstMaxWidthForm[2]{}
\providecommand*\mciteBstWouldAddEndPuncttrue
  {\def\EndOfBibitem{\unskip.}}
\providecommand*\mciteBstWouldAddEndPunctfalse
  {\let\EndOfBibitem\relax}
\providecommand*\mciteSetBstMidEndSepPunct[3]{}
\providecommand*\mciteSetBstSublistLabelBeginEnd[3]{}
\providecommand*\EndOfBibitem{}
\mciteSetBstSublistMode{f}
\mciteSetBstMaxWidthForm{subitem}{(\alph{mcitesubitemcount})}
\mciteSetBstSublistLabelBeginEnd
  {\mcitemaxwidthsubitemform\space}
  {\relax}
  {\relax}

\bibitem[Hewson \latin{et~al.}(2011)Hewson, Kapur, and
  Gaskell]{Hewson-2011-ID152}
Hewson,~R.; Kapur,~N.; Gaskell,~P. A two-scale model for discrete cell gravure
  roll caoting. \emph{Chemical Engineering Science} \textbf{2011}, \emph{66},
  3666--3674\relax
\mciteBstWouldAddEndPuncttrue
\mciteSetBstMidEndSepPunct{\mcitedefaultmidpunct}
{\mcitedefaultendpunct}{\mcitedefaultseppunct}\relax
\EndOfBibitem
\bibitem[Schneider \latin{et~al.}(2019)Schneider, Losio, Nüesch, and
  Heier]{Schneider-2019-ID308}
Schneider,~R.; Losio,~P.~A.; Nüesch,~F.~A.; Heier,~J. Gravure printed
  Ag/conductive polymer electrodes and simulation of their electrical
  properties. \emph{The International Journal of Advanced Manufacturing
  Technology} \textbf{2019}, \relax
\mciteBstWouldAddEndPunctfalse
\mciteSetBstMidEndSepPunct{\mcitedefaultmidpunct}
{}{\mcitedefaultseppunct}\relax
\EndOfBibitem
\bibitem[Jilesen \latin{et~al.}(2015)Jilesen, Gaylard, Spruss, Kuthada, and
  Wiedemann]{Jilesen-2015-ID310}
Jilesen,~J.; Gaylard,~A.; Spruss,~I.; Kuthada,~T.; Wiedemann,~J. Advances in
  Modelling A-Pillar Water Overflow. \emph{{SAE} Technical Paper Series}
  \textbf{2015}, \emph{1}\relax
\mciteBstWouldAddEndPuncttrue
\mciteSetBstMidEndSepPunct{\mcitedefaultmidpunct}
{\mcitedefaultendpunct}{\mcitedefaultseppunct}\relax
\EndOfBibitem
\bibitem[Dianat \latin{et~al.}(2017)Dianat, Skarysz, and
  Garmory]{Dianat-2017-ID311}
Dianat,~M.; Skarysz,~M.; Garmory,~A. A Coupled Level Set and Volume of Fluid
  method for automotive exterior water management applications.
  \emph{International Journal of Multiphase Flow} \textbf{2017}, \emph{91},
  19--38\relax
\mciteBstWouldAddEndPuncttrue
\mciteSetBstMidEndSepPunct{\mcitedefaultmidpunct}
{\mcitedefaultendpunct}{\mcitedefaultseppunct}\relax
\EndOfBibitem
\bibitem[Princen(1969)]{Princen-1969-ID288}
Princen,~H. Capillary Phenomena in Assemblies of Parallel Cylinders.
  \emph{Journal of Colloid and Interface Science} \textbf{1969}, \emph{30},
  69--75\relax
\mciteBstWouldAddEndPuncttrue
\mciteSetBstMidEndSepPunct{\mcitedefaultmidpunct}
{\mcitedefaultendpunct}{\mcitedefaultseppunct}\relax
\EndOfBibitem
\bibitem[Clanet and Qu\'er\'e(2002)Clanet, and Qu\'er\'e]{Clanet-2002-ID206}
Clanet,~C.; Qu\'er\'e,~D. Onset of menisci. \emph{Journal of Fluid Mechanics}
  \textbf{2002}, \emph{460}\relax
\mciteBstWouldAddEndPuncttrue
\mciteSetBstMidEndSepPunct{\mcitedefaultmidpunct}
{\mcitedefaultendpunct}{\mcitedefaultseppunct}\relax
\EndOfBibitem
\bibitem[Ponomarenko \latin{et~al.}(2011)Ponomarenko, Qu\'er\'e, and
  Clanet]{Ponomarenko-2011-ID202}
Ponomarenko,~A.; Qu\'er\'e,~D.; Clanet,~C. A universal law for capillary rise
  in corners. \emph{Journal of Fluid Mechanics} \textbf{2011}, \emph{666},
  146--154\relax
\mciteBstWouldAddEndPuncttrue
\mciteSetBstMidEndSepPunct{\mcitedefaultmidpunct}
{\mcitedefaultendpunct}{\mcitedefaultseppunct}\relax
\EndOfBibitem
\bibitem[Hsieh and Yu(1992)Hsieh, and Yu]{Hsieh-1992-ID303}
Hsieh,~Y.-L.; Yu,~B. Liquid Wetting, Transport, and Retention Properties of
  Fibrous Assemblies. \emph{Textile Research Journal} \textbf{1992}, \emph{62},
  677--685\relax
\mciteBstWouldAddEndPuncttrue
\mciteSetBstMidEndSepPunct{\mcitedefaultmidpunct}
{\mcitedefaultendpunct}{\mcitedefaultseppunct}\relax
\EndOfBibitem
\bibitem[Qu\'er\'e \latin{et~al.}(1988)Qu\'er\'e, Di~Meglio, and
  Brochard-Wyart]{Quere-1988-ID297}
Qu\'er\'e,~D.; Di~Meglio,~J.-M.; Brochard-Wyart,~F. Wetting of fibers : theory
  and experiments. \emph{Revue de Physique Appliquée} \textbf{1988},
  \emph{23}, 1023--1030\relax
\mciteBstWouldAddEndPuncttrue
\mciteSetBstMidEndSepPunct{\mcitedefaultmidpunct}
{\mcitedefaultendpunct}{\mcitedefaultseppunct}\relax
\EndOfBibitem
\bibitem[Fuentes \latin{et~al.}(2011)Fuentes, Tran, Dupont-Gillain,
  Vanderlinden, De~Feyter, Van~Vuure, and Verpoest]{Fuentes-2011-ID299}
Fuentes,~C.; Tran,~L.; Dupont-Gillain,~C.; Vanderlinden,~W.; De~Feyter,~S.;
  Van~Vuure,~A.; Verpoest,~I. Wetting behaviour and surface properties of
  technical bamboo fibres. \emph{Colloids and Surfaces A} \textbf{2011},
  \emph{380}, 89--99\relax
\mciteBstWouldAddEndPuncttrue
\mciteSetBstMidEndSepPunct{\mcitedefaultmidpunct}
{\mcitedefaultendpunct}{\mcitedefaultseppunct}\relax
\EndOfBibitem
\bibitem[Aranberri-Askargorta \latin{et~al.}(2003)Aranberri-Askargorta, Lampke,
  and Bismarck]{Aranberri-Askargorta-2003-ID298}
Aranberri-Askargorta,~I.; Lampke,~T.; Bismarck,~A. Wetting behavior of flax
  fibers as reinforcement for polypropylene. \emph{Journal of Colloid and
  Interface Science} \textbf{2003}, \emph{263}, 580--589\relax
\mciteBstWouldAddEndPuncttrue
\mciteSetBstMidEndSepPunct{\mcitedefaultmidpunct}
{\mcitedefaultendpunct}{\mcitedefaultseppunct}\relax
\EndOfBibitem
\bibitem[Hsieh \latin{et~al.}(1996)Hsieh, Miller, and
  Thompson]{Hsieh-1996-ID301}
Hsieh,~Y.-L.; Miller,~A.; Thompson,~J. Wetting, Pore Structure, and Liquid
  Retention of Hydrolyzed Polyester Fabrics. \emph{Textile Research Journal}
  \textbf{1996}, \emph{66}, 1--10\relax
\mciteBstWouldAddEndPuncttrue
\mciteSetBstMidEndSepPunct{\mcitedefaultmidpunct}
{\mcitedefaultendpunct}{\mcitedefaultseppunct}\relax
\EndOfBibitem
\bibitem[Kim and Hsieh(2001)Kim, and Hsieh]{Kim-2001-ID300}
Kim,~C.; Hsieh,~Y.-L. Wetting and absorbency of nonionic surfactant solutions
  on cotton fabrics. \emph{Colloids and Surfaces A} \textbf{2001},
  \emph{187-188}, 385--397\relax
\mciteBstWouldAddEndPuncttrue
\mciteSetBstMidEndSepPunct{\mcitedefaultmidpunct}
{\mcitedefaultendpunct}{\mcitedefaultseppunct}\relax
\EndOfBibitem
\bibitem[Yanılmaz and Kalaoğlu(2012)Yanılmaz, and
  Kalaoğlu]{Yanilmaz-2012-ID302}
Yanılmaz,~M.; Kalaoğlu,~F. Investigation of wicking, wetting and drying
  properties of acrylic knitted fabrics. \emph{Textile Research Journal}
  \textbf{2012}, \emph{82}, 820--831\relax
\mciteBstWouldAddEndPuncttrue
\mciteSetBstMidEndSepPunct{\mcitedefaultmidpunct}
{\mcitedefaultendpunct}{\mcitedefaultseppunct}\relax
\EndOfBibitem
\bibitem[Bico \latin{et~al.}(2001)Bico, Tordeux, and Quéré]{Bico-2001-ID209}
Bico,~J.; Tordeux,~C.; Quéré,~D. Rough wetting. \emph{Europhysics Letters}
  \textbf{2001}, \emph{55}, 214--220\relax
\mciteBstWouldAddEndPuncttrue
\mciteSetBstMidEndSepPunct{\mcitedefaultmidpunct}
{\mcitedefaultendpunct}{\mcitedefaultseppunct}\relax
\EndOfBibitem
\bibitem[Palasantzas and De~Hosson(2001)Palasantzas, and
  De~Hosson]{Palasantzas-2001-ID307}
Palasantzas,~G.; De~Hosson,~J. T.~M. Wetting on rough surfaces. \emph{Acta
  Materialia} \textbf{2001}, \emph{49}, 3533--3538\relax
\mciteBstWouldAddEndPuncttrue
\mciteSetBstMidEndSepPunct{\mcitedefaultmidpunct}
{\mcitedefaultendpunct}{\mcitedefaultseppunct}\relax
\EndOfBibitem
\bibitem[Boreyko \latin{et~al.}(2011)Boreyko, Baker, Poley, and
  Chen]{Boreyko-2011-ID210}
Boreyko,~J.~B.; Baker,~C.~H.; Poley,~C.~R.; Chen,~C.-H. Wetting and dewetting
  transitions on hierarchical superhydrophobic surfaces. \emph{Langmuir}
  \textbf{2011}, \emph{27}, 7502--7509\relax
\mciteBstWouldAddEndPuncttrue
\mciteSetBstMidEndSepPunct{\mcitedefaultmidpunct}
{\mcitedefaultendpunct}{\mcitedefaultseppunct}\relax
\EndOfBibitem
\bibitem[Papadopoulos \latin{et~al.}(2012)Papadopoulos, Deng, Mammen, Drotlef,
  Battagliarin, Li, Müllen, Landfester, del Campo, Butt, and
  Vollmer]{Papadopoulos-2012-ID208}
Papadopoulos,~P.; Deng,~X.; Mammen,~L.; Drotlef,~D.-M.; Battagliarin,~G.;
  Li,~C.; Müllen,~K.; Landfester,~K.; del Campo,~A.; Butt,~H.-J.; Vollmer,~D.
  Wetting on the microscale: shape of a liquid drop on a microstructured
  surface at different length scales. \emph{Langmuir} \textbf{2012}, \emph{28},
  8392--8398\relax
\mciteBstWouldAddEndPuncttrue
\mciteSetBstMidEndSepPunct{\mcitedefaultmidpunct}
{\mcitedefaultendpunct}{\mcitedefaultseppunct}\relax
\EndOfBibitem
\bibitem[Giacomello \latin{et~al.}(2016)Giacomello, Schimmele, and
  Dietrich]{Giacomello-2016-ID280}
Giacomello,~A.; Schimmele,~L.; Dietrich,~S. Wetting hysteresis induced by
  nanodefects. \emph{Proceedings of the National Academy of Sciences of the
  United States of America} \textbf{2016}, \emph{113}, E262--71\relax
\mciteBstWouldAddEndPuncttrue
\mciteSetBstMidEndSepPunct{\mcitedefaultmidpunct}
{\mcitedefaultendpunct}{\mcitedefaultseppunct}\relax
\EndOfBibitem
\bibitem[Rye \latin{et~al.}(1996)Rye, Yost, and Mann]{Rye-1996-ID305}
Rye,~R.~R.; Yost,~F.~G.; Mann,~J. A.~J. Wetting Kinetics in Surface Capillary
  Grooves. \emph{Langmuir} \textbf{1996}, \emph{12}, 4625--4627\relax
\mciteBstWouldAddEndPuncttrue
\mciteSetBstMidEndSepPunct{\mcitedefaultmidpunct}
{\mcitedefaultendpunct}{\mcitedefaultseppunct}\relax
\EndOfBibitem
\bibitem[Yost \latin{et~al.}(1997)Yost, Rye, and Mann]{Yost-1997-ID304}
Yost,~F.~G.; Rye,~R.~R.; Mann,~J. A.~J. Solder wetting kinetics in narrow
  V-grooves. \emph{Acta Materialia} \textbf{1997}, \emph{45}, 5337--5345\relax
\mciteBstWouldAddEndPuncttrue
\mciteSetBstMidEndSepPunct{\mcitedefaultmidpunct}
{\mcitedefaultendpunct}{\mcitedefaultseppunct}\relax
\EndOfBibitem
\bibitem[Seemann \latin{et~al.}(2005)Seemann, Brinkmann, Kramer, Lange, and
  Lipowsky]{Seemann-2005-ID306}
Seemann,~R.; Brinkmann,~M.; Kramer,~E.~J.; Lange,~F.~F.; Lipowsky,~R. Wetting
  morphologies at microstructured surfaces. \emph{Proceedings of the National
  Academy of Sciences of the United States of America} \textbf{2005},
  \emph{102}, 1848--1852\relax
\mciteBstWouldAddEndPuncttrue
\mciteSetBstMidEndSepPunct{\mcitedefaultmidpunct}
{\mcitedefaultendpunct}{\mcitedefaultseppunct}\relax
\EndOfBibitem
\bibitem[Herminghaus \latin{et~al.}(2008)Herminghaus, Brinkmann, and
  Seemann]{Herminghaus-2008-ID3}
Herminghaus,~S.; Brinkmann,~M.; Seemann,~R. Wetting and Dewetting of Complex
  Surface Geometries. \emph{Annual Review of Materials Research} \textbf{2008},
  \emph{38}, 101--121\relax
\mciteBstWouldAddEndPuncttrue
\mciteSetBstMidEndSepPunct{\mcitedefaultmidpunct}
{\mcitedefaultendpunct}{\mcitedefaultseppunct}\relax
\EndOfBibitem
\bibitem[Deng \latin{et~al.}(2014)Deng, Tang, Zeng, Yang, and
  Shao]{Deng-2014-ID271}
Deng,~D.; Tang,~Y.; Zeng,~J.; Yang,~S.; Shao,~H. Characterization of capillary
  rise dynamics in parallel micro V-grooves. \emph{International Journal of
  Heat and Mass Transfer} \textbf{2014}, \emph{77}, 311--320\relax
\mciteBstWouldAddEndPuncttrue
\mciteSetBstMidEndSepPunct{\mcitedefaultmidpunct}
{\mcitedefaultendpunct}{\mcitedefaultseppunct}\relax
\EndOfBibitem
\bibitem[O'Brien \latin{et~al.}(1968)O'Brien, Craig, and
  Peyton]{O'Brien-1968-ID293}
O'Brien,~W.~J.; Craig,~R.~G.; Peyton,~F.~A. Capillary Penetration between
  Dissimilar Solids. \emph{Journal of Colloid and Interface Science}
  \textbf{1968}, \emph{26}, 500--508\relax
\mciteBstWouldAddEndPuncttrue
\mciteSetBstMidEndSepPunct{\mcitedefaultmidpunct}
{\mcitedefaultendpunct}{\mcitedefaultseppunct}\relax
\EndOfBibitem
\bibitem[Tang and Tang(1994)Tang, and Tang]{Tang-1994-ID240}
Tang,~L.-H.; Tang,~Y. Capillary Rise in Tubes with Sharp Grooves. \emph{Journal
  de Physique {II}} \textbf{1994}, \relax
\mciteBstWouldAddEndPunctfalse
\mciteSetBstMidEndSepPunct{\mcitedefaultmidpunct}
{}{\mcitedefaultseppunct}\relax
\EndOfBibitem
\bibitem[Concus and Finn(1969)Concus, and Finn]{Concus-1969-ID294}
Concus,~P.; Finn,~R. On the behavior of a capillary surface in a wedge.
  \emph{Proceedings of the National Academy of Sciences of the United States of
  America} \textbf{1969}, \emph{63}, 292--299\relax
\mciteBstWouldAddEndPuncttrue
\mciteSetBstMidEndSepPunct{\mcitedefaultmidpunct}
{\mcitedefaultendpunct}{\mcitedefaultseppunct}\relax
\EndOfBibitem
\bibitem[Thammanna~Gurumurthy \latin{et~al.}(2018)Thammanna~Gurumurthy,
  Rettenmaier, Roisman, Tropea, and Garoff]{ThammannaGurumurthy-2018-ID232}
Thammanna~Gurumurthy,~V.; Rettenmaier,~D.; Roisman,~I.~V.; Tropea,~C.;
  Garoff,~S. Computations of spontaneous rise of a rivulet in a corner of a
  vertical square capillary. \emph{Colloids and Surfaces A: Physicochemical and
  Engineering Aspects} \textbf{2018}, \emph{544}, 118--126\relax
\mciteBstWouldAddEndPuncttrue
\mciteSetBstMidEndSepPunct{\mcitedefaultmidpunct}
{\mcitedefaultendpunct}{\mcitedefaultseppunct}\relax
\EndOfBibitem
\bibitem[Thammanna~Gurumurthy \latin{et~al.}(2018)Thammanna~Gurumurthy,
  Roisman, Tropea, and Garoff]{ThammannaGurumurthy-2018-ID230}
Thammanna~Gurumurthy,~V.; Roisman,~I.~V.; Tropea,~C.; Garoff,~S. Spontaneous
  rise in open rectangular channels under gravity. \emph{Journal of Colloid and
  Interface Science} \textbf{2018}, \emph{527}, 151--158\relax
\mciteBstWouldAddEndPuncttrue
\mciteSetBstMidEndSepPunct{\mcitedefaultmidpunct}
{\mcitedefaultendpunct}{\mcitedefaultseppunct}\relax
\EndOfBibitem
\bibitem[Brakke(1992)]{Brakke-1992-ID295}
Brakke,~K.~A. The Surface Evolver. \emph{Experimental Mathematics}
  \textbf{1992}, \emph{1}, 141--165\relax
\mciteBstWouldAddEndPuncttrue
\mciteSetBstMidEndSepPunct{\mcitedefaultmidpunct}
{\mcitedefaultendpunct}{\mcitedefaultseppunct}\relax
\EndOfBibitem
\bibitem[Bullard and Garboczi(2009)Bullard, and Garboczi]{Bullard-2009-ID289}
Bullard,~J.~W.; Garboczi,~E.~J. Capillary rise between planar surfaces.
  \emph{Physical Review: E} \textbf{2009}, \emph{79}, 011604\relax
\mciteBstWouldAddEndPuncttrue
\mciteSetBstMidEndSepPunct{\mcitedefaultmidpunct}
{\mcitedefaultendpunct}{\mcitedefaultseppunct}\relax
\EndOfBibitem
\end{mcitethebibliography}

\end{document}